\documentclass[sigconf]{acmart}

\usepackage{xcolor}
\usepackage{ifthen}
\usepackage{CJKutf8}
\newboolean{cameraready}
\setboolean{cameraready}{true} 

\ifthenelse{\boolean{cameraready}}
  {\colorlet{maincolor}{black}} 
  {\colorlet{maincolor}{blue}}  

\begin{document}

\title{AI, Climate, and Regulation: From Data Centers to the AI Act}

\author{Kai Ebert}
\authornote{Both authors contributed equally to this research.}
\email{kebert@europa-uni.de}
\orcid{0009-0002-5250-0587}
\affiliation{%
  \institution{European University Viadrina}
  \city{Frankfurt (Oder)}
  \country{Germany}
}

\author{Nicolas Alder}
\authornotemark[1]
\affiliation{%
  \institution{Hasso Plattner Institute}
  \city{Potsdam}
  \country{Germany}}
\email{nicolas.alder@hpi.de}

\author{Ralf Herbrich}
\affiliation{%
  \institution{Hasso Plattner Institute}
  \city{Potsdam}
  \country{Germany}}
\email{ralf.herbrich@hpi.de}

\author{Philipp Hacker}
\affiliation{%
 \institution{European University Viadrina}
 \city{Frankfurt (Oder)}
 \country{Germany}}
\email{hacker@europa-uni.de}


\begin{abstract}
We live in a world that is experiencing an unprecedented boom of AI applications that increasingly penetrate all sectors of private and public life, from education, media, medicine, and mobility to the industrial and professional workspace. As this world is simultaneously grappling with climate change, the climate effects and environmental implications of the development and use of AI have become an important subject of public and academic debate. In this paper, we aim to provide guidance on the climate-related regulation for data centers and AI specifically, and discuss how to operationalize these requirements. We also highlight challenges and room for improvement, and make a number of policy proposals to this end. In particular, we propose a specific interpretation of the AI Act to bring reporting on the previously unaddressed energy consumption from AI inferences back into the scope. We also find that the AI Act fails to address indirect greenhouse gas emissions from AI applications. Furthermore, for the purpose of energy consumption reporting, we compare levels of measurement within data centers and recommend measurement at the cumulative server level. We also argue for an interpretation of the AI Act that includes environmental concerns in the mandatory risk assessments (sustainability risk assessment, SIA), and provide guidance on its operationalization. The EU data center regulation proves to be a good first step but requires further development by including binding renewable energy and efficiency targets for data centers. Overall, we make twelve concrete policy proposals, in four main areas: Energy and Environmental Reporting Obligations; Legal and Regulatory Clarifications; Transparency and Accountability Mechanisms; and Future Far-Reaching Measures beyond Transparency.
\end{abstract}

\maketitle

\section{Introduction}\label{sec:introduction}
The environmental consequences of artificial intelligence (AI) are becoming ever more apparent as large models are increasingly trained and deployed across society. This inherently intertwines the digital transformation with questions of climate change, particularly concerning the energy and water consumption of AI models. These features are attracting attention from both the public and academia (see below, 2.). 
Concerns are growing that the supply of renewable energy may not keep up with their increasing demand triggered by the AI scaling race \cite{alder2024ai} in which the EU is poised to join the U.S. and China under the EU Commission's new AI Continent Action Plan \cite{eu2025aicontinent}. \textcolor{maincolor}{Scholars and activists, in turn, are increasingly critiquing and situating digital technology within the framework of environmental justice \cite{Kazansky2022,sotolongo2023defining,yang2025path,Rakova2023,Mozilla2024}. Work in political economy has shed a light on the extractive nature of AI-based socio-technical systems, extracting and, at times, exploiting, both natural and human resources \cite{Keskin2021,Kasy2024,EkbiaNardi2017,Crawford2021atlas,mills2024algorithms}, often particularly from marginalized communities \cite{Perrigo2023,Noble2018} and regions \cite{Li2023load,yang2021aitocracy}. Both the quest for performance and the funding logic behind AI force providers to scale them to a point where this trajectory may push against planetary limits \cite{Brevini2021,Crawford2021atlas,brevini2024ecopolitical,bender2021dangers}}.

\textcolor{maincolor}{Regulatory frameworks are beginning to address these challenges, too. While the new U.S. administration scraps environmental rules and generally deregulates, the EU Artificial Intelligence Act (AI Act) becomes applicable as the world’s most comprehensive attempt at direct AI regulation. It features environmental protection as one of its core goals and contains dedicated sustainability rules, which also apply to US and other non-EU providers offering models in the EU. Similarly, the Digital Services Act (DSA) compels Very Large Online Platforms and Very Large Online Search Engines to thoroughly assess and mitigate systemic risks, which is of particular relevance for hybrid platforms increasingly integrating AI. As scholars have pointed out, this includes climate and sustainability risks \cite{buri2024platform}, and intersects partially with obligations under the AI Act \cite{Hacker2024aiact,Helberger2023chatgpt}. In parallel, the Energy Efficiency Directive (EU) 2023/1791 introduces transparency obligations for data centers, a critical infrastructure supporting AI. However, these regulatory instruments were developed independently, leading to fragmentation and gaps. The AI Act originally pursued climate-related objectives especially in the European Parliament's position of June 2023, but over successive drafts and the trilogue negotiations, its environmental provisions were diluted, particularly with regard to energy consumption from AI inference and risk assessments. Meanwhile, the EU’s data center regulation remains incomplete, lacking binding efficiency and renewable energy targets. These shortcomings threaten to undermine the most promising legislative avenues for tackling AI's climate effects. The EU's recent push for massive investments in computing infrastructure under the AI Continent Action Plan, including the establishment of AI Gigafactories \cite{eu2025aicontinent}, is a late but necessary impetus for technological and strategic autonomy but must not endanger the Union's climate goals under the Green New Deal.}

\textcolor{maincolor}{Against this background, this paper makes three key contributions. First, we offer the first thorough analysis of the AI Act's final version from an environmental perspective. Unlike prior scholarship, which has examined these areas separately, we also scrutinize the interplay between data center regulation and direct AI regulation. AI sustainability is not governed by a single legal instrument but rather emerges from the interaction of multiple frameworks and actors. This fragmented approach creates legal uncertainty and regulatory loopholes — notably the omission of AI inference from reporting obligations and the failure to mandate Sustainability Impact Assessments (SIA). We then propose a novel interpretation of the AI Act to include AI inferences in the reporting obligations of general-purpose AI providers, and to incorporate sustainability considerations into the mandatory risk assessment framework to effectuate the goal of environmental protection enshrined in Art. 1 and Recitals 1, 2 and 8.}

\textcolor{maincolor}{Second, we examine the practical challenges of implementing AI-related environmental obligations and provide technical recommendations. The crux here is to operationalize existing rules and make climate reporting work by developing metrics that are both meaningful and workable in practice.} 

 \textcolor{maincolor}{Third, we identify areas where the AI Act and related regulatory frameworks require future amendments. In light of upcoming evaluation rounds, we advocate for closing loopholes and reinstating environmental obligations that were weakened during the legislative process. While the current political momentum seems to favor deregulation, the climate emergency persists. Therefore, it is crucial to improve the best legal tools still available, and to advocate for a nuanced legal framework to bring the twin transitions of AI and climate together. Our twelve policy proposals address four key areas: Energy and Environmental Reporting Obligations; Legal and Regulatory Clarifications; Transparency and Accountability Mechanisms; and Future Far-Reaching Measures beyond Transparency.}

\section{Related Work}
\textcolor{maincolor}{Due to the interdisciplinary nature of sustainable AI, related works originate from three main fields---computer science; humanities and policy; and legal research. In recent years, the computer science and AI community has become increasingly aware of the environmental effects of AI, which led to both policy-oriented \cite{schwartz2020green,freitag2021real,chien2021good,bender2021dangers,knowles2021acm,taddeo2021artificial,kaack2022aligning,knowles2022our,luccioni2024environment,luccioni2024light,luccioni2025jevons} and technical contributions, particularly on the environmental effects of data centers operations and AI-related computing \cite{oecd2022aiimpact,guidi2024environmental,EPRI2024}, studies for tracking \cite{dayarathna2015data,luccioni2024power,alder2024energy,luccioni2023counting,luccioni2023estimating,wu2022sustainable,dodge2022measuring,patterson2021carbon,zhou2020hulk} and techniques for reducing the emissions of data centers and AI models \cite{beloglazov2011taxonomy,koomey2021pitfalls,guler2021framework,ludvigsen2022carbonfootprint,kneese2024carbon}. The water consumption of AI training has also been a concern \cite{li2023making,zuccon2023beyond}.}
\textcolor{maincolor}{Broader scholarship and policy work includes ethical \cite{coeckelbergh2021ai,floridi2021philosophical}, informational \cite{luccioni2024light,stein2020artificial,pagallo2022environmental,devries2023growing}, and social perspectives \cite{bakhtiarifard2025climate}. Numerous contributions also analyze and evaluate the various means in which AI may be used to mitigate, and adapt to, climate change \cite{stein2020artificial,cowls2023ai,kelly2022ethical,saheb2022artificial,hamdan2024ai,chen2023artificial,rolnick2022tackling,kumari2023application}, as also acknowledged in the AI Act (Recitals 4 and 142).
}
\textcolor{maincolor}{However, from a legal perspective, the problem of sustainable AI remains underexplored. Existing contributions date from before the AI Act's final version \cite{pagallo2022environmental}, which differs significantly from previous proposals (see below, 5.2.) or do not engage with its provisions in detail \cite{hacker2024sustainable}. Other, complimentary work focuses uniquely on the DSA \cite{buri2024platform} or data center regulation \cite{Commins2025}. 
}

\section{Technical Background}
From a technical perspective, it is important to distinguish between (pre-)training, fine-tuning, and inference. Training refers to the process of initially adjusting a model's parameters or weights to fit the data. This process is highly compute-intensive and typically requires a significant amount of energy \cite{luccioni2024power, luccioni2023estimating, wu2022sustainable, kaack2022aligning}. LLM performance strongly depends on the model scale (number of parameters), which in turn requires more training data \cite{kaplan2020scaling, rae2021scaling}, and hence more resources. Such scaling can even be expected with algorithmically advanced models, such as DeepSeek R1 \cite{Amodei2025DeepSeek}. 

Fine-tuning aims at adjusting a pretrained model to fit more specific data in order to perform better at specific tasks. 
It involves a significantly smaller amount of training data and compute budget. The specific procedure for successful fine-tuning is very model-, data-, and task-specific, and rather empirical. Therefore, the computational and energy-intensity can vary significantly for this step. For example, as highlighted by Luccioni et al. \cite{luccioni2024power}, the energy usage for fine-tuning the Bloomz-7B required 7,571 kWh compared to 51,686 kWh for the entire training process, adding another 15 \% to the initial consumption. 
Inference, in turn, refers to the procedure of generating a prediction from a trained model. Typically, an individual inference consumes little energy compared to training; but there are many more inference than training events. 

Most of these training, fine-tuning and inference computations are conducted in data centers. The power usage effectiveness (PUE) metric reflects the energy efficiency of a data center. It indicates the ratio of the total energy needed by a data center, including components such as cooling, to the energy used solely by computational devices. A PUE of 1.0 would imply ideal efficiency, meaning that the data center uses only the energy necessary to power the computational devices. 
The average data center PUE in 2023 was 1.58 globally\cite{statista2025pue} and 1.6 in the EU \cite{jrc2023datacentres}.

\section{Regulation of Data Centers}
Data centers run all kinds of operations, such as cloud computing, crypto currencies and the Internet at large. Recently, AI training and inference have experienced massive growth. Broad data center regulation, therefore, indirectly governs the environmental effects of AI and constitutes the backdrop against which specific AI regulation must be viewed. Taking a look at the requirements for data centers can help to find a coherent and effective interpretation of the requirements in the AI Act as it builds on already available data and established methodology.



\subsection{EU Data Collection and Reporting Obligations for Data Centers} 
\label{EU Data Collection and Reporting Obligations for Data Centers}
In the EU, data collection and reporting obligations for data centers were established by two recent legal acts, Art. 12 of the recast Energy Efficiency Directive EU/2023/1791 of September 13, 2023 ("EED"), and the Commission Delegated Regulation EU/2024/1364 of March 14, 2024 ("Delegated Regulation"). The new rules apply to all data centers in the EU with a power demand of the installed information technology (IT) of at least 500kW, which includes small-sized data centers. Data center operators are required to collect, make publicly available and report to a EU database information that is deemed relevant for the sustainability assessment of the data centers and the industry as a whole. The reporting is mandated on an annual basis. 

The required data includes energy consumption, power utilization, temperature set points, waste heat utilisation, water usage and use of renewable energy (EED, Annex VII(c)). Notably, while the EED focuses on energy and power, the reporting of water usage is a significant step forward as both energy and water consumption have raised concerns in AI settings \cite{li2023making,luccioni2024power}. In addition, the Water Framework Directive can be harnessed to limit the overall amount water a data center may consume, and also control for any potential loss of water quality \cite{hacker2024sustainable}.

The Delegated Regulation provides specific key performance indicators and methodology. Most notable is the requirement to measure and report the energy consumption of the installed information technology. Following the standard-methodology for the calculation of PUE,\footnote{The Delegated Regulation refers to the European standard CEN/CENELEC EN 50600-4-2.} the energy consumption must be measured at the uninterruptible power system (UPS) or, if not existent, at the power distribution unit (PDU) or at another point specified by the data center (see Delegated Regulation, Annex II(1)(e); see also the appendix, Figure 1 as Categories 1-3.

The data centers must report to the EU database directly or via a national reporting scheme, if such a scheme is implemented by the Member State. From the reported data the Commission calculates the data center sustainability indicators which are made publicly available on an aggregate level. They include the power usage effectiveness (PUE), the water usage effectiveness (WUE)\footnote{WUE measures the amount of water used in the data center's cooling and other operations relative to the energy consumed by the IT stack.}, the energy reuse factor (ERF)\footnote{ERF represents the percentage of energy that is reused from a data center's waste energy, showing the efficiency of energy recovery.}, and the renewable energy factor (REF)\footnote{REF indicates the proportion of a data center's energy that comes from renewable sources.}.


\subsection{Energy Management Systems and Energy Audits} \label{Energy Management Systems and Energy Audits}
The EED also requires that Member States mandate companies with an average annual energy consumption of more than 10 TJ to conduct an energy audit at least every four years and those with a consumption of more than 85 TJ to implement an energy management system including regular energy audits (Art. 11 EED). This would also apply to operators of data centers. The Directive sets up certain minimum criteria for energy audits (Annex VI EED) and refers to the relevant international or European standards (Recital 80 EED). The legal minimum criteria, however, do not dictate how energy consumption should be measured. 

\subsection{German 2023 Energy Efficiency Act} \label{German 2023 Energy Efficiency Act}
In Germany, the Energy Efficiency Act of 8 Nov 2023 implements the EED and establishes a national reporting scheme and additional requirements, including specific efficiency and renewable energy targets for data centers. The Act broadens the scope of the reporting obligation to include even smaller data centers, upwards of 300 kW (Sec. 13). It also expands the duty to set up an energy management system to data centers and operators of ICT---i.e., customers of colocation data centers---of more than 50 kW (Sec. 12). Most importantly, it sets targets on energy efficiency and renewable energy use, requiring data centers to reach a PUE factor between 1.5 and 1.2 and an ERF of 10\% to 20 \% depending on their age (Sec. 11), and to run on 50 \% renewable energy, increasing that factor to 100\% by 1 Jan 2027 (Sec. 11). Lastly, it requires data center operators to inform their customers on an annual basis on the energy consumption directly attributable to them (Sec. 15).


\subsection{Regulation Outside the EU}
Outside the EU, a number of countries is taking the lead by implementing concrete PUE goals, such as Singapore, Japan, China, and Australia, while the U.S. as one of the main contributors in the AI race shows a nuanced picture. However, it should be noted that some of these measures are limited to procurement policies or mere political commitments where it is unclear if there will be any penalty-enabled enforcement like in the German Energy Efficiency Act. A direct comparison of PUE target values should, therefore, be made with this caveat in mind.

Under the 2024 Green Data Center Roadmap, Singapore aims to reach a PUE of 1.3 or less within 10 years \cite{singapore2024gdcr}. Japan, with its 2022 Energy Conservation Act, requires data center operators to take efficiency measures to reach a target PUE of 1.4 by 2030 \cite{businessnetwork2024}. Meanwhile, in China, according to Li et al., policies issued from 2013 onward introduced efficiency targets, gradually decreasing the required PUE from 1.5 in 2013 to 1.3 in 2021 \cite{Li2023China}. However, the authors also admit that most recent data still shows a significant implementation gap with the actual PUE of most data centers ranging between 1.4 and 2.0. As part of Australia’s 2023 Net Zero in Government Operations Strategy, data center services procured by the government must demonstrate a five-star energy rating under the national NABERS scheme, including a PUE of 1.4 or lower. This requirement applies to procurement from members of the distinguished Data Centre Panel, and from 1 July 2025, it will extend to other companies and state-operated data centers \cite{aus2023netzero}.

In the U.S., in the absence of federal efficiency requirements, the only direct regulation for data center efficiency is the California Green Building Action Plan 2015 implementing Executive Order B-18-12. Data centers that exceed a PUE of 1.5 are required to reduce their PUE by a minimum of 10 percent per year until they achieve a 1.5 or lower PUE. However, similar to the Australian action, this plan is only limited to state data centers \cite{california2012green,MM14-09}. 

Another noteworthy strain of regulation is the climate reporting set forth in the SEC climate disclosure rules, and, on a state level, the California Climate Corporate Data Accountability Act, also known as SB 253. While the California Act goes further in including private companies and scope 3 indirect emissions, it is limited to companies with USD 1 B in annual revenue. For the SEC rules as well as SB 253 the focus is on the companies' total emissions, therefore direct attribution to AI or computing energy consumption will be difficult. 

Relating to AI more specifically, although not limited to data centers, is a bill for an AI Environmental Impacts Act that was introduced in the U.S. Senate by Senator Edward J. Markey (D-MA) on 1 Feb 2024 \cite{uscongress2024s3732}. The bill was referred to the Committee on Commerce, Science and Transportation, and has not yet been voted upon. Under the new administration, it is unlikely that the bill will pass Congress. However, even if enacted, it would not contain any significant hard regulation. The bill primarily mandates studies, stakeholder consultations, and voluntary reporting on AI's environmental impacts without imposing significant regulatory obligations.


\subsection{Discussion and Interim Conclusion on Data Center Regulation}
The regulation of data centers in the context of AI’s environmental impact, particularly regarding energy and water consumption, presents both advantages and shortcomings. The increasing growth of AI-related activities, such as training and inference, places significant pressure on the environmental footprint of data centers. While the EU has implemented generic data center regulations, such as those outlined in the Energy Efficiency Directive and the Delegated Regulation, these rules also indirectly govern the environmental impact of AI by imposing reporting and data collection requirements. Notably, these regulations require the reporting of both energy and water consumption, a critical aspect given the rising concerns over resource use in AI applications.

One of the key strengths of the EU’s approach is the establishment of specific reporting obligations. Data center operators must collect and publicly report energy consumption, power utilization, water usage, waste heat utilization, and the use of renewable energy. These measures help create transparency and provide a foundation for future efficiency improvements. Additionally, the German implementation of the EED goes beyond mere reporting by setting specific targets for energy efficiency and renewable energy use in data centers, as well as requiring smaller data centers to comply. Germany’s approach might serve as a potential blueprint for broader EU regulation, particularly mandating data center operators to inform customers about their direct energy consumption, an essential factor for optimizing AI-related energy use.

However, the regulations also present several shortcomings. While data collection and reporting obligations are useful, the absence of binding efficiency and renewable energy targets at the EU level is a major limitation. Although the Commission is expected to propose further legislative measures by 2025, the current lack of enforceable standards means that data centers could continue to consume vast amounts of energy and water without significant reductions in their environmental impact. Moreover, while Germany has introduced stricter targets, these do not extend to all Member States, potentially leading to a fragmented regulatory environment across the EU.

In contrast, the situation outside the EU shows a very diverse landscape, ranging from government procurement to industry wide political commitments and concrete legal requirements with varying implementation success.



\section{AI Act: Gaps and Interpretation Challenges}
The AI Act applies further down in the value chain, targeting entities that develop and deploy AI systems. After introducing the key terminology used by the Act, we discuss the major climate-related obligations contained in it, with a particular focus on transparency, risk assessment and mitigation. As our analysis shows, the Act makes steps in the right direction, but falls short of a comprehensive regime for tackling climate-related risks of AI models.

\subsection{Terminology of the AI Act: AI, Providers, and Deployers}
The AI Act only applies if an AI model or AI system is used by a specifically designated entity, such as a provider or deployer. The activity also needs to relate to the EU, typically either because the system is offered to persons in the EU or because its output is used in the EU (Art. 2(1) and Recital 22 AI Act). An AI system is defined as a machine-based system 'designed to operate with varying levels of autonomy and that may exhibit adaptiveness after deployment, and that, for explicit or implicit objectives, infers, from the input it receives, how to generate outputs such as predictions, content, recommendations, or decisions that can influence physical or virtual environments' (Art. 3(1)).

Key actors in the AI value chain include providers and deployers. Providers develop or market AI models or systems (Art. 3(3)) while deployers use existing AI systems in a professional capacity (Art. 3(4)). Note that other operationally relevant actors, such as cloud service providers, data providers, data center operators, or other intermediaries, are not directly subject to AI Act rules.

The Act pays specific attention to general-purpose AI (GPAI) models, e.g., large language models (LLMs) like GPT-4, that are defined by the wide range of possible uses. Those with the most advanced capabilities, and therefore higher risk of negative effects, are labeled GPAI models with systemic risk (Art. 3(63)-(65)). An AI system that integrates a GPAI model is a general-purpose AI system (Art. 3(66)), e.g., ChatGPT. AI systems used in specific settings subject to product safety regulation, e.g., medical devices, and areas particularly sensitive for public safety or fundamental rights, e.g., law enforcement, education, employment, or credit scoring, are called high-risk AI (HRAI) systems. These systems are subject to other rules, concerning training data, documentation, human in the loop, and performance, inter alia (Art. 9-16). When GPAI systems are used in high-risk applications, rules may apply cumulatively (Recital 85, 97).

The AI Act contains energy- and climate-related transparency and risk management obligations, primarily for \emph{providers} of AI systems or models. Under certain conditions (e.g., fine-tuning), however, deployers become providers, which triggers much more onerous duties, also with respect to climate impacts (Art. 25(1), Recital 109). Hence, there is a palpable incentive for companies to avoid provider status.

Concerning HRAI systems, pursuant to Art. 25(1), deployers can become providers if they market an existing HRAI system under their name or trademark, substantially modify an HRAI system, or change the intended purpose of a non-high-risk AI system such that it falls under one of the high-risk sectors (e.g., employment; life- or health insurance; education). Specifically, if a deployer uses a GPAI system, such as ChatGPT, in a high-risk setting, for example, by harnessing it for resume screening in hiring, they automatically assume the responsibilities of a provider of an HRAI system (Art. 25(1)(c)). 

An existing HRAI system can be altered. If the modification is substantial, the modifying entity becomes a new provider (Article 25(1)(b)). It is unclear, however, what precisely constitutes a "substantial modification" that would trigger provider status for the modifying entity. A formal interpretation suggests that any change to the model could be deemed substantial, based on Art. 3(23), which defines substantial modification as any unplanned change affecting compliance or intended purpose. On the other hand, a---more convincing---material interpretation would claim that provider status should only apply if the modification increases the model's risk in a nonnegligible way. This aligns with the AI Act's focus on risk mitigation and also the language in Article 3(23) according to which any significant modification must affect compliance with the high-risk rules of the Act (or change the purpose). 

For GPAI models, Recitals 97, 109, and 111 suggest that any entity performing even minimal fine-tuning of a GPAI model is automatically classified as the provider of a new model, including all associated responsibilities \cite{alder2024ai}. For minor modifications this is disproportionate, even if Recital 109 states that obligations are limited to the fine-tuning itself \cite{alder2024ai}. A solution could be an analogy with Article 25(1)(b) so that provider status would only be triggered by a \textit{substantial} modification made through fine-tuning, rather than any minor changes \cite{alder2024ai}.

This is particularly important for SMEs using powerful models (GPAI models with systemic risk): if they become providers, they need to comply with the much more stringent obligations in Article 55 concerning evaluation and risk management, including environmental risk.
\subsection{Legislative History and Intent}
\textcolor{maincolor}{In the original Commission proposal of April 2021, the AI Act only included minimal references to environmental effects, mostly as a potential benefit to climate change mitigation (see above, 2.), and in the section on voluntary commitments \cite{pagallo2022environmental}. The General Approach adopted by the Council in December 2022 continued this narrow framing. However, the momentum changed, particularly with the position of the European Parliament (EP), published in June 2023. The suggested EP sustainability amendments to the AI Act introduce environmental principles for AI development, preferential funding for eco-friendly AI, mandatory energy and resource consumption reporting, sustainability risk assessments (SIAs) for foundation models, and Commission-led guidance on environmental impact measurement and review \cite{hacker2024sustainable}. During the final trilogue negotiations, several of these amendments were dropped (e.g., the principle; explicit SIAs). However, simultaneously, the impact of fundamental rights was strengthened, and environmental protection listed as one such right that must be considered in a range of provisions \cite{hacker2024sustainable,pagallo2022environmental}. This is supported by the mention of environmental protection as a key policy goal of the AI Act in Article 1 and Recitals 1, 2 and 8.}

\subsection{Transparency}
The first climate-related obligation in the AI Act concerns transparency and reporting, however not with respect to data centers, but concerning providers of high-risk and GPAI models \cite{alder2024ai}. Yet, the rules come with significant ambiguities and loopholes. In the following, we detail the six most important ones.

1) For high-risk AI systems, pursuant to Article 11(1) providers must document the computational resources utilized during the development, training, testing, and validation stages, as outlined in Annex IV(2) which does, however, not include the energy consumption directly \cite{alder2024ai}. This omission hinders the capacity to evaluate and compare the environmental impact of such systems accurately. As a result, the environmental footprint must be estimated indirectly based on the recorded computational resources.

2) The AI Act requires providers of general-purpose AI models to meet transparency obligations. Article 53(1)(a) mandates that providers maintain up-to-date technical documentation, including the details outlined in Annex XI. This annex requires the reporting of energy consumption, whether known or estimated, while estimates may be based on the computational resources used. However, this requirement exhibits a significant gap as it only covers the energy used during the model’s development phase, but leaves out the inference phase \cite{alder2024ai}. Recent research has shown that energy consumption during inference often exceeds that of the development phase very significantly over time \cite{luccioni2024power, wu2022sustainable}.

To bridge this gap, we propose an alternative interpretation, based on the information provided to downstream actors (see also \cite{alder2024ai}). As a result of the AI Act, providers must supply downstream developers and relevant authorities with technical information on how to incorporate GPAI models into AI systems (Articles 53(1)(a) and (b), together with Annex XI and Annex XII). While energy consumption is not explicitly referenced in these provisions, they could reasonably be understood to include details about the computational resources required for inference because downstream providers need this information to properly allocate resources, e.g., purchasing computing power for operating the model. As a side-effect, this information could be used to calculate energy consumption for individual inferences. However, to calculate the overall energy consumption for inferences, the total number of inferences would need to be available as well. While the typical energy costs of inferences should therefore be mandatorily disclosed by providers, deployers should be compelled to divulge information on the number of inferences, as well, e.g., on an aggregate monthly level.

3) Open-source (OS) GPAI models are largely excluded from transparency requirements unless they present a systemic risk, as outlined in Articles 2(12) and 53(2) \cite{alder2024ai}. The idea is that OS models, by definition, already disclose certain types of information. Hence, Recital 102 lists information on parameters, including weights, model architecture, and model usage as a prerequisite for systems to be considered OS. However, it does not mandate the disclosure of energy consumption. This results in an unwarrented lack of visibility concerning the environmental impact and energy usage of these models.

4) Where the Act does mandate the disclosure of energy consumption, this information is restricted to authorities and is not accessible to downstream providers (unless the proposed interpretation from 2) is applied) or the general public, due to confidentiality clauses in Articles 21(3), 53(7), and 78(1) \cite{alder2024ai}. The limited availability of this data significantly reduces transparency and accountability, theryby weakening the potential for public oversight and market responses.

5) The AI Act fails to address the greenhouse gas (GHG) emissions generated by AI applications, for instance in sectors like oil and gas exploration \cite{kaack2022aligning, alder2024ai}. For example, a recent investigation has revealed Microsoft's aggressive pitch of its AI models to ExxonMobile to optimize fossil fuel exploration \cite{hao2024microsoft}. Leaving such applications with significant climate effects out of scope creates a notable reporting gap.

6) The Act also neglects to address the substantial water consumption, a key concern in data center operations \cite{alder2024ai}. While the Energy Efficiency Directive mandates reporting of water consumption for data centers, the AI Act requires no such reporting specifically for AI---as it does for energy use---nor does it cover operations outside the EU.

\subsection{Environmental Risk Assessment and Mitigation}
In tackling the climate effects of AI and ICT more generally, it is arguably crucial to move beyond mere transparency provisions towards more substantive goals and obligations. Indeed, the AI Act does contain some language to this effect. For providers of GPAI models with systemic risk and providers of HRAI systems, the Act mandates risk assessment and mitigation (Art. 55(1)(b) and Art. 9). We argue that these measures should also consider environmental risks\textcolor{maincolor}{, in keeping with the normative goals of the AI Act listed in Article 1 and Recitals 1, 2 and 8.
}


Crucially, both provisions relate to risks of the AI model or system for fundamental rights which, within the AI Act, must be interpreted as including environmental risks \cite{hackerberz2025}. In Art. 1(1) and Recital 1, the purpose of the AI Act is defined as protecting health, safety, fundamental rights enshrined in the Charter, including democracy, the rule of law and environmental protection. However, in the doctrine of the Charter of Fundamental Rights of the European Union (the Charter), environmental protection (Art. 37 of the Charta) is merely an objective rule, not a fundamental right \cite{morgera_duran_2022}. Democracy and the rule of law are not enshrined in a particular Article of the Charter but serve as guiding principles that permeate the Charter and all of the fundamental rights. Accordingly, they (only) find a mention in the preamble of the Charter.

In our view, the European legislator did not err on the doctrinal classification of democracy, the rule of law and environmental protection in the Charter by including these principles. While it is conceivable that a legislative error may have occurred in miscategorising the objective rule in Art. 37 of the Charta, that seems very unlikely for the basic principles merely expressed in the preamble. Instead, the more convincing interpretation is that, for the purposes of the AI Act, the legislator meant for these principles to be included whenever the Act refers to fundamental rights. It is, besides, not uncommon for a legal term to have different meanings in different legal settings.

\textcolor{maincolor}{Thus, it should not be dispositive that the explicit SIA contained in the EP position (see 5.2) was not included in the final version. Rather,} the provisions on risk assessment and mitigation form a centerpiece of the protective measures included in the AI Act. Without including environmental risks in the risk assessment and mitigation, in effect, not much consideration for environmental risks would remain and certainly not the "high level of protection" that Art. 1(1) mandates.

As a caveat, it should be noted that, while the AI Act requires climate risk assessment and mitigation, no detailed reporting is mandated. For HRAI systems the documentation must comprise a detailed \textit{description} of the risk management system, but not the risk assessment or its results (Annex IV(5)). Providers of GPAI models with systemic risk may rely on codes of practice or European harmonized standards, both of which are not yet available, or alternative means of compliance for assessment by the Commission (Art. 55(2)). If these will entail a detailed reporting of the risk assessment, and to whom, is yet to be determined.

\subsection{Discussion and Interim Conclusion on the AI Act}
Overall, while the AI Act introduces valuable steps toward addressing climate-related concerns in AI development and deployment, it falls short of establishing a comprehensive framework for mitigating the environmental risks posed by AI systems. Key provisions on transparency and risk management for high-risk AI and general-purpose AI systems make some progress in requiring documentation of computational resources and energy consumption, but significant gaps remain. For example, the Act does not mandate the disclosure of energy consumption during the inference phase, a crucial omission given the long-term environmental impact of AI applications. Moreover, transparency measures are restricted to authorities, limiting broader accountability and public scrutiny.

Additionally, while the Act imposes risk assessment and mitigation obligations on providers of HRAI systems and GPAI models with systemic risk, these provisions lack sufficient emphasis on environmental factors. Although environmental protection is included in the Act's objectives, its practical integration into risk management remains unclear and no detailed reporting on mitigation efforts concerning environmental risks is currently required. Without stronger enforcement and clearer guidance, particularly on the inclusion of energy consumption and other environmental impacts in risk assessments, the Act's potential to address the growing climate-related risks of AI systems will remain limited.

\section{Operationalizing the Requirements}

\subsection{Model Providers and their Access to Infrastructure}
Leading model providers leverage extensive supercomputer networks equipped with high-performance GPUs. This group is most impacted by energy consumption reporting. We assume that they possess privileged access or considerable influence over the utilized infrastructure given their financial power or strategic value. One can distinguish between companies that develop closed models (e.g., OpenAI) and those that create open-source models (e.g., Meta). 

Startups and smaller companies are expected to either directly deploy such models, use available API services of leading model providers, or fine-tune existing models. Although these companies do not typically train large models themselves, they must report energy consumption for fine-tuning. Under our proposed interpretation, however, only if the fine-tuning of a model leads to a substantial modification. 

\subsection{Levels for Measuring and Estimating Energy Consumption}
There are several levels within a data center based on which energy consumption may be measured or estimated \cite{alder2024ai}. These include (1) the data center level, (2) the cumulative server level, (3) the GPU-Level and other hardware within a server and (4) various other levels. In this section, we outline each level along with their benefits, drawbacks, and estimation methods for when measurement is unavailable. 

\subsubsection{Data Center Level} 
On the data center level, the power required to operate the entire data center is measured, including both the direct power consumption of computing equipment and the additional overhead for cooling and maintaining the data center. This approach provides the most extensive and complete figures since it represents the actual energy usage, but also assumes that a data center is exclusively utilized for the pre-training by the model provider. It encourages the selection of an efficient data center. Additionally, data centers have average PUE values of 1.58, so this overhead makes up a significant portion of the energy consumption. On the other hand, the power usage resulting solely from the model's architecture, the quantity of training data, the efficiency of the implementation, and experimental settings is very important but is somewhat skewed by the efficiency of the data center. 

If data-center level power consumption measurement is not available, using the PUE factor for estimation is deemed appropriate. To calculate total energy usage, the PUE factor is multiplied by the raw computational power consumption measured or estimated at the cumulative server or rack level (see below). This might be reasonable, if only parts of a data center are utilized and only measurements closer to the ICT equipment are available.

\subsubsection{Cumulative Server Level} 
\begin{figure}[h]
\centering
  \includegraphics[width=1\linewidth]{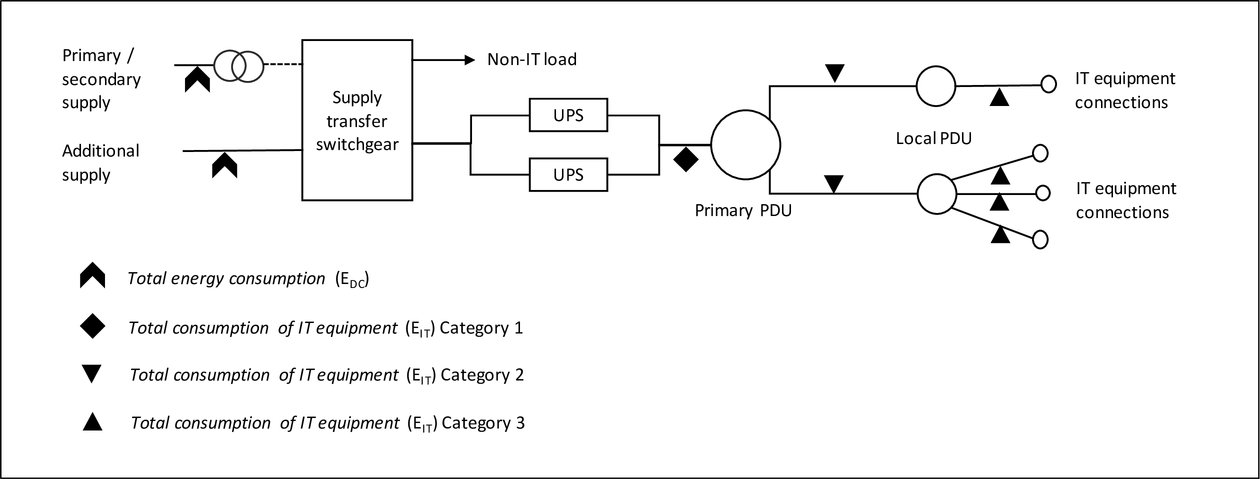}
\caption{Measuring points for IT equipment energy consumption, from Annex II of the Delegated Regulation EU/2024/1364 of March 14, 2024}
    \label{fig:DD-measuring-points}
\end{figure}

A large-scale model is trained across many servers in a distributed manner. Each server includes GPUs responsible for the primary computation. To accurately monitor the power consumption over time, a local power distribution unit (PDU), capable of measuring the provided power, is attached to each server. Aggregating these measurements yields a highly precise figure of the total energy consumption attributable to the model's computations. Instead of aggregating local PDUs, the usage of primary PDUs or uninterruptible power supply (UPS) systems already measuring at the rack level or even many racks is also suitable (See also Appendix, Figure \ref{fig:DD-measuring-points}), as long as the measurements precisely match fully the utilized hardware resources by the model providers. The goal is to include all ICT-related power consumption but exclude data center specific efficiency properties.

The upside of this method is its high accuracy, highly correlating with model size and structure, data quantity, and hardware-aware software implementation. It is  widely recognized in the industry for assessing power consumption in data centers. According to a 2023 Green Grid industry survey \cite{greengrid2023survey}, 66\% of data centers can track power demand at least on rack-level. Roughly one third of overall data centers are already able to collect average utilization and power demand data for individual servers and storage equipment and match this data to their IT equipment inventory. We assume that the data centers that are used for training by large model providers are already able to track this information given its high cost relevance. However, a significant number of data centers do not track power demand yet. The surveyed data center professionals estimate to require between 3-6 months (Europe: 15\%, Global: 19\%), 1 year (Europe: 29\%, Global: 28\%), 2 years (Europe: 12\%, Global: 10\%), 3 years (Europe: 4\%, Global: 4\%) or more than 4 years (Europe: 11\%, Global: 8\%) to implement adequate power collection abilities. For European data centers these numbers stand in (surprising) contrast to the obligation under the EED and Delegated Regulation to provide energy consumption data by Sep 15, 2024 (Art. 3(1) and Annex II(1)(e) Delegated Regulation). Either a substantial number of the participants was unaware of the obligations; or thereby acknowledged their inability to comply in time; or the question was too narrow, being pointed at individual servers, while the EED also allows measurement at the higher-level UPS.

An estimation is possible when GPU hours and the hardware used are known. By multiplying a GPU-hour with the peak utilization power consumption specified by the manufacturer, one can estimate the upper limit of energy consumption at the server level. This upper bound is typically higher than the actual consumption, as GPU utilization rarely reaches its theoretical peak due to other resource constraints. Higher GPU efficiency means that the same operations are completed in less time, reducing the overhead from non-utilization-dependent hardware power consumption.

\subsubsection{GPU-Level and other hardware within a server} 
The measurement may be based on the energy usage of particular components as determined by on-chip sensors. Nvidia GPUs and certain CPUs already provide straightforward power consumption monitoring, therefore estimations are usually not necessary. However, despite GPU power consumption being a significant factor and its usage correlating with the total power usage, it substantially under-represents the actual energy consumption since it measures just a single component. CPU power usage is a relatively minor factor in consumption. Most other server components cannot be measured. We advocate against using GPU-level or other component-based power consumption tracking for overall energy measurements. 

\subsubsection{Other levels} 
Other measurement levels, such as Workload, Cloud Instance, or Virtual Machine, involve high complexity and numerous assumptions, resulting in a lack of standardized measurement or estimation methods with considerable uncertainty. 
We advise against using these levels for power consumption tracking.

\subsubsection{Interim Conclusion on the Level of Measurement}
In our analysis, we argue that energy consumption should be measured and reported at the cumulative server level (see also \cite{alder2024ai}). This approach captures the total computation-related power usage and is better suited to help providers optimize their AI models and algorithms for energy efficiency. Additionally, the PUE factor of each data center, which is reported and published by the data center operator under the Energy Efficiency Directive (EU) 2023/1791 and Delegated Regulation (EU) 2024/1364, provides a useful estimate of overall energy consumption \cite{alder2024ai}. With these two figures, it is possible to distinguish between model-specific power usage (server-level computation) and the data center’s efficiency, offering a clearer picture of the total energy investment \cite{alder2024ai}. Although the EED mandates that PUE factors must be available for data centers situated within the EU, the responsibility of reporting these factors should also fall on the model provider. Specifically, model providers utilizing data center facilities outside the EU should not be granted an exemption.

Estimates of server-level power consumption should be based on peak utilization figures provided by the hardware manufacturer (e.g., Nvidia) \cite{alder2024ai}. Still, it is important to consider advancements in research. Interpretation of the legal requirements could accommodate justifiable alternative assumptions that may not require peak utilization figures. For instance, tracking GPU utilization through interfaces like those provided by Nvidia and referencing hardware benchmarks based on specific GPU utilization rates could serve as a basis for such assumptions.

\subsection{Measurement or Estimation}
Although actual measurement is more onerous, it also yields more precise results for energy consumption reporting. Major model providers are likely to already measure power consumption as it is a primary cost factor and highly linked to computational power. Despite the availability of power consumption data, companies may be tempted to use estimated values to protect sensitive information. This practice should be restricted by legally prioritizing measurements over estimations in Annex XI Section 1(2)(e). 

The same reasoning applies to smaller entities relying on cloud computing services for fine-tuning, for example such offered by Amazon Web Services or Microsoft Azure. Fine-tuning is crucial for employing foundational models in task-specific applications and tailoring them to specific datasets. The higher the expense of initial model trainings, the stronger the incentive to perform fine-tuning rather than retraining the model. Therefore, this is the type of adaptation that most businesses will focus on to effectively integrate large AI models into practical products. 

Cloud platforms still lack client-oriented power consumption reporting as part of their products. It is essential that ordinary companies with limited resources do not face obstacles in fulfilling their reporting obligations only because their cloud computing providers do not offer this data. The access rights enshrined in Art. 25(2) AI Act can help if the deploying entity becomes a provider---but providers may invoke trade secrets and IP rights to dilute this obligation (Art. 25(5) AI Act). In our view, the interests of downstream actors in reporting correct figures should generally trump the secrecy interests of upstream providers, in this case. Under the current setup, companies could always resort to computation-based estimations for their reporting. However, in order to gain the most accurate data, cloud platforms should be incentivized to provide energy consumption data to their clients. Complementing Art. 25(2) AI Act, the obligation established in Sec. 15 of the German Energy Efficiency Act, requiring data centers to inform customers on their attributable annual energy consumption, could serve as a blueprint. Notably, such a law could also apply to cloud service providers based outside the EU, with the caveat, however, that such an extraterritorial application oftentimes lacks enforcement capabilities.

\subsection{Sustainability Impact Assessments}
The operationalization of sustainability impact assessments (SIAs) within the risk assessments required under the AI Act involves integrating environmental considerations into the existing risk management frameworks that high-risk AI model providers and GPAI providers must follow. Much like data protection or algorithmic impact assessments, SIAs would serve as a practical tool for embedding climate considerations into the development and deployment of AI systems. Importantly, these assessments should not be limited to high-risk AI models but should also apply to all AI systems, regardless of the associated risk to health or safety. This is because the carbon footprint of AI models is often unrelated to their classification as high or low risk under the Act. Therefore, an SIA could ensure that environmental impacts are considered across the entire AI landscape.

The SIA should involve evaluating various models and design choices during the development process, comparing them not only on their performance but also on their estimated environmental impact. For instance, developers would need to assess whether a simpler model, like linear regression or even a non-AI model, could achieve similar results with a smaller carbon footprint compared to more complex models like deep learning \cite{kaack2022aligning,hacker2024sustainable}. Similarly, the decision to use large, pre-trained models or training new, narrow models (almost) from scratch should factor in the potential climate benefits. By using existing tools to measure the carbon impact of AI models, developers would be required to opt for the more environmentally sustainable option when performance is comparable. To effectively implement sustainability impact assessments, providers would need to establish standardized methodologies for measuring the environmental effects of AI models, particularly energy and water usage during training and inference phases.



Taking one step back, the concept of "data protection by design" should be expanded to include “sustainability by design,” under which developers should actively seek to reduce the contribution of ICT, including AI, to climate change \cite{hacker2024sustainable}. At both the technical and organizational levels, this would involve adopting all feasible measures to limit environmental impact, a shift that has already been applied in other industries through consumption practices and product design. These approaches are also gaining traction in supply chain management and other sectors in pursuit of corporate Environmental, Social, and Governance (ESG) objectives. By drawing on these existing practices, sustainability by design could become a core principle guiding the regulation of the ICT and AI sectors.

\section{Policy Proposals}
Although the AI Act attempts to address climate concerns through various reporting obligations, these measures largely lack consistency and clarity. We identify twelve policy recommendations that should be integrated into the evaluation report due in August 2028 (Article 111(6)), as well as any interpretive guidelines from the AI Office and other agencies, and in reviews and potential textual revisions prior to that date (see also \cite{alder2024ai}). These measures can be grouped in four categories: Energy and Environmental Reporting Obligations; Legal and Regulatory Clarifications; Transparency and Accountability Mechanisms; and Future Far-Reaching Measures beyond Transparency.

\subsection{Energy and Environmental Reporting Obligations}

The current AI Act overlooks key environmental factors related to AI systems \cite{luccioni2024environment}. The following proposals aim to ensure comprehensive energy and environmental accountability for AI systems, both for development and inference phases.

\begin{itemize}
    \item \textbf{Inclusion of Energy Consumption From Inferences}: We propose an interpretation that would at least bring energy consumption for single inferences back into the scope (see also \cite{alder2024ai}). Adoption of this interpretation by courts, authorities and industry, however, is uncertain. Therefore, both single and overall inferences should be included as a reporting category in Annex XI and Annex XII (vis-à-vis authorities and downstream providers) through delegated acts from the Commission (Articles 53(5) and (6)) and future recommendations from the AI Office. 
    
    \item \textbf{Indirect Emissions and Water Consumption Reporting}: The Act currently omits indirect emissions from AI applications (e.g., those used for oil and gas exploration \cite{kaack2022aligning}) and water consumption \cite{li2023making}. Reporting should include: Providers reporting water usage, and deployers reporting application-related emissions, allowing for estimates where precise measurements are impossible, particularly when dealing with future scenarios.
    
    \item \textbf{Energy Reporting at the Cumulative Server Level}: Energy consumption should be reported at the cumulative server level (see also \cite{alder2024ai}). In this endeavor, estimations may be used only when direct measurements are unavailable. These principles should guide the development of technical standards under Article 40 of the AI Act, as well as the potential implementation of the Global Digital Compact on a global scale.
\end{itemize}


\subsection{Legal and Regulatory Clarifications on AI Models and Providers}

Ambiguities in the AI Act regarding provider responsibilities for AI models need to be clarified to ensure that entities know when and what to comply with. The following proposals seek ensure that companies understand their obligations, that open-source models are on par with closed models, and that the regulatory framework sets effective incentives at the source (data centers) to mitigate the environmental impact of AI model development and deployment.

\begin{itemize}
    \item \textbf{Clarifying Provider Status for Model Modifications}: The current definition of provider status needs to focus on substantial modifications to AI models, specifically those that involve changing the model's weights. Reporting obligations and the change-of-provider threshold should be tied to computational costs incurred during significant modifications, with a minimum computational cost threshold ensuring that only energy-intensive changes trigger additional reporting \cite{alder2024ai}.
    
    \item \textbf{Elimination of the Open-Source Exemption}: The open-source exemption from reporting obligations should be removed, as making parts of a model public does not justify exclusion from environmental accountability \cite{alder2024ai}. Open-source models can have significant energy implications and should adhere to the same reporting standards as proprietary models.
    
    \item \textbf{Energy Efficiency and Renewable Energy Targets for Data Centers}: The EED and Delegated Regulation should be amended in three regards, following the German Energy Efficiency Act: they should set binding efficiency targets (PUE) and renewable energy targets for data centers; and include an information obligation for data center operators and cloud service providers vis-à-vis their customers regarding the attributable energy consumption. These measures should be included in the 2025 evaluation report. This would both increase transparency and facilitate more sustainable decisions by market participants.
\end{itemize}


\subsection{Transparency and Accountability Mechanisms}

To promote public trust and accountability in AI’s environmental impact, the following measures are recommended:

\begin{itemize}
    \item \textbf{Public Access to Climate-Related Disclosures}: All climate-related disclosures, including energy consumption, should be made accessible to the general public \cite{alder2024ai}. If only aggregate data are shared, trade secrets can be protected while allowing for public scrutiny by analysts, academics, and NGOs. Public transparency would drive market pressure, reputational incentives, and public accountability; it would, arguably, encourage companies to reduce their environmental impact.
    
    \item \textbf{Energy Reporting for High-Risk AI (HRAI)}: For High-Risk AI (HRAI) systems, the Act should require the disclosure of energy consumption rather than computational resources to eliminate inaccuracies and enhance comparability. 
    
    \item \textbf{Environmental Risk Assessments}: Providers should be required to include environmental risks in their risk assessments. The language in Art. 1(1) and Recital 1, as well as Art. 9 and 55 AI Act should be clarified to reflect this. This will ensure that environmental impacts are systematically evaluated alongside other risks during AI system development and deployment.
\end{itemize}


\subsection{Beyond Transparency}

In addition to the existing proposals, more far-reaching measures that go beyond reporting and transparency may need to be considered in the future, particularly as AI’s energy demands grow. The following proposals focus on reducing AI’s strain on energy resources and fostering more sustainable energy practices:

\begin{itemize}
    \item \textbf{Restrictions on AI Training and Inference During Peak Hours}: One future measure could involve limiting AI training and certain inference tasks (e.g., those for recreation-related purposes) to non-peak energy demand hours \cite{hacker2024sustainable}. This would reduce the strain on energy grids during peak times and make AI energy usage more manageable and sustainable.
    
    \item \textbf{Obligations for Data Centers and AI Companies to Create Renewable Energy Sources}: AI companies and data centers should be required to create new renewable energy sources to offset any excess energy demand caused by their operations. This measure could help ensure that renewable energy, which is a finite resource, is not monopolized by AI and data centers, leaving enough clean energy for the rest of the industry. Clear definitions and guidelines would need to be established to define which types of energy consumption would be covered by this 'new source requirement'.
    
    \item \textbf{Tradable Energy Budgets for AI}: Another possible long-term measure could involve implementing tradable energy budgets for AI, similar to the EU Emissions Trading System (EU ETS) \cite{hacker2024sustainable}. Companies that consume particularly large amounts of energy would have a capped energy budget for AI operations and could trade these credits. This would create a market-based approach to managing and reducing AI-related energy consumption. It would, arguably, encourage companies to optimize their energy usage and invest in energy efficiency.
\end{itemize}



 \section{Objections and Solutions}

\textcolor{maincolor}{Implementing the proposed policy recommendations presents both practical and political challenges, particularly given the EU’s broader regulatory landscape and the evolving international context. This section outlines four key objections to our proposals and sketches solutions.
First, a central priority of the new European Commission is the simplification of the digital acquis (‘fitness check’) \cite{mlex2025euaiact}. There is concern that adding further reporting requirements and regulatory obligations could create administrative burdens that counteract this goal. However, several of our proposals, particularly standardized energy reporting metrics and clearer environmental risk assessments, align with the EU’s push for regulatory clarity and streamlining. Rather than adding complexity, these measures would, arguably, make compliance more predictable and feasible.}

\textcolor{maincolor}{Second, the shifting international AI governance landscape creates uncertainty regarding the feasibility of ambitious environmental rules. There is concern that EU regulatory efforts might face resistance from international AI providers or risk competitive disadvantages. While acknowledging a trend toward deregulation in some corners of the world, we argue that this reinforces the need for robust EU policy leadership. The AI Act’s extraterritorial scope already compels global AI providers to comply with EU regulations when operating in the European market. By introducing clear, enforceable environmental obligations, the EU can set a global regulatory precedent, and shape international norms despite external pressures. Moreover, some of our proposals - such as market-based mechanisms for energy efficiency - offer flexible compliance pathways that may be more politically acceptable in a deregulatory global environment.}

\textcolor{maincolor}{Third, the AI industry, particularly large-scale AI providers and data center operators, may resist additional regulatory obligations, citing economic and technological constraints. There is also concern that lack of enforcement mechanisms could render some obligations ineffective. Hence, our proposals emphasize practical enforcement strategies, including gradual implementation, industry incentives, and transparency-driven accountability. For example: public access to climate-related disclosures can create reputational incentives for compliance, even in the absence of strict enforcement; energy efficiency targets for data centers can be phased in over time for a realistic transition without abrupt regulatory shocks; clarifications on provider status and reporting obligations offer legal certainty to companies.
Fourth, some of our more ambitious proposals - such as tradable energy budgets for AI or restrictions on AI training during peak hours - may be viewed as far-reaching or difficult to implement in the near term. While we recognize that not all proposals will be immediately adopted, it is crucial to articulate ambitious policy options at this juncture. The AI regulatory landscape is evolving, and incremental steps toward long-term measures can be integrated into future evaluation rounds of the AI Act, the EED, and related frameworks. As with the EU Emissions Trading System (ETS), initial skepticism toward market-based environmental regulations can give way to gradual acceptance as regulatory mechanisms mature.
Given the current political momentum and upcoming AI Act evaluation rounds, there is a narrow but critical window to influence environmental AI governance. While not all recommendations will be immediately adopted, presenting a structured, well-argued policy package ensures that sustainability concerns remain central to AI regulation in the EU and beyond - rather than being sidelined in the face of deregulation.}

\section{Conclusion}
AI systems, many of them containing energy-intensive generative AI components, are increasingly integrated into economic and societal processes. Importantly, as AI intersects with robotics, writing code and steering devices, it is poised to merge decisively into industrial processes, proliferating its deployment, but also its climate effects. Currently, this intersection between the AI and the green transition sits at a regulatory blind spot, inadequately addressed in current regulation. 
In this paper, we take a closer look at sustainability-related data center regulation and the sustainability provisions in the AI Act. While they present a good first step, they too often lack ambition, clarity and consistency, or present significant challenges in implementation. To counter these shortcomings, we provide interpretations of the AI Act in line with its purpose of environmental protection, provide guidance on operationalizing the requirements, and make twelve concrete proposals, grouped into four areas: Energy and Environmental Reporting Obligations; Legal and Regulatory Clarifications; Transparency and Accountability; and Future Far-Reaching Measures beyond Transparency.

\bibliographystyle{ACM-Reference-Format}
\bibliography{bibfile}
\newpage
\appendix
\section{Measuring Points for IT Equipment in Data Centers}

\newpage
\clearpage


\section{Policy Proposals Overview}
\begin{table}[h]
\centering
\renewcommand{\arraystretch}{1.3}  
\captionsetup{skip=10pt}  
\caption{Summary of Policy Proposals for AI and Environmental Impact}
\begin{tabular}{p{4cm}p{10cm}l}
\toprule

\textbf{Area} & \textbf{Policy Proposal and Description} \\
\cmidrule(r){1-2}
\textbf{Energy and Environmental Reporting} 
    & \textbf{Energy consumption from inferences}: Include energy consumption from both single and cumulative inferences in reporting. \\
    & \textbf{Indirect emissions and water consumption}: Mandate reporting of indirect emissions and water use in AI applications. \\
    & \textbf{Cumulative server energy reporting}: Require energy consumption to be measured and reported at the cumulative server level. \\ \cmidrule(r){1-2}
    
\textbf{Legal and Regulatory Clarifications} 
    & \textbf{Clarifying provider status for AI modifications}: Limit provider status to substantial AI model modifications. \\
    & \textbf{Elimination of open-source exemption}: Remove the exemption that allows open-source models to bypass reporting obligations. \\
    & \textbf{Renewable energy and efficiency targets for data centers}: Introduce binding energy efficiency and renewable energy targets for data centers. \\ \cmidrule(r){1-2}
    
\textbf{Transparency and Accountability} 
    & \textbf{Public access to climate-related disclosures}: Ensure that climate-related disclosures are accessible to the public, not just authorities. \\
    & \textbf{Energy reporting for High-Risk AI}: Replace computational resource reporting with direct energy consumption disclosures for High-Risk AI systems. \\
    & \textbf{Inclusion of environmental risks in assessments}: Clarify that environmental risks must be part of risk assessments for AI providers. \\ \cmidrule(r){1-2}
    
\textbf{Future Far-Reaching Measures} 
    & \textbf{Restrictions during peak energy hours}: Limit AI training and non-essential inference tasks to non-peak energy demand periods. \\
    & \textbf{Obligation to create renewable energy sources}: Require AI companies and data centers to develop renewable energy sources to offset excess energy use. \\
    & \textbf{Tradable energy budgets for AI}: Introduce a market-based system with tradable energy budgets for AI operations, similar to the EU ETS. \\ \bottomrule

\end{tabular}

\end{table}

\end{document}